\documentclass[10pt]{article}

\usepackage{amsmath}
\usepackage{pstricks}
\usepackage{pst-node}
\usepackage{pst-3dplot}
\usepackage{fullpage}
\usepackage{url}
\usepackage[bf]{caption}
\usepackage[hidelinks]{hyperref}

\def\BibTeX{{\rm B\kern-.05em{\sc i\kern-.025em b}\kern-.08em
    T\kern-.1667em\lower.7ex\hbox{E}\kern-.125emX}}

\title{A systolic update scheme to overcome memory bandwidth limitations in GPU-accelerated FDTD simulations}

\author{
  Jesse Lu\\
  SPINS Photonics Inc\\
  \href{mailto:jesselu@spinsphotonics.com}{jesselu@spinsphotonics.com}
  \and
  David Qu\\
  SPINS Photonics Inc\\
  \href{mailto:davidqu@spinsphotonics.com}{davidqu@spinsphotonics.com}
  \and
  Jim Qu\\
  SPINS Photonics Inc\\
  \href{mailto:jimqu@spinsphotonics.com}{jimqu@spinsphotonics.com}
  \and
  Ryan Fong\\
  SPINS Photonics Inc\\
  \href{mailto:ryanfong@spinsphotonics.com}{ryanfong@spinsphotonics.com}
  \and
  Geun Ho Ahn\\
  Stanford University\\
  \href{mailto:gahn@stanford.edu}{gahn@stanford.edu}
  \and
  Jelena Vu\v ckovi\'c\\
  Stanford University\\
  \href{mailto:jela@stanford.edu}{jela@stanford.edu}
}

\begin{document}

\maketitle

\begin{abstract}

The exponential growth of artificial intelligence has fueled
  the development of high-bandwidth photonic interconnect fabrics
  as a critical component of modern AI supercomputers~\cite{jouppi2023tpu}.
As the demand for ever-increasing AI compute and connectivity continues to grow,
  the need for high-throughput photonic simulation engines
  to accelerate and even revolutionize photonic design and verification workflows
  will become an increasingly indispensable capability for the integrated photonics industry.
Unfortunately,
  the mainstay and workhorse of photonic simulation algorithms,
  the finite-difference time-domain (FDTD) method~\cite{yee1966numerical},
  because it is a memory-intensive, but computationally-lightweight algorithm,
  is fundamentally misaligned with modern computational platforms,
  which are equipped to deal with compute-intensive workloads instead.
This paper introduces a systolic update scheme for the FDTD method,
  which circumvents this mismatch by reducing the need for global synchronization
  while also relegating the need to access global memory
  to the case where boundary values between neighboring subdomains only.
We demonstrate a practical implementation of our scheme
  as applied to the full three-dimensional FDTD algorithm 
  that achieves a performance of $\sim 0.15$ trillion cell updates per second (TCUPS)
  on a single Nvidia H100 GPU.
Our work paves the way for the increasingly efficient, cost-effective, and high-throughput
  photonic simulation engines needed to continue powering the AI era.

\end{abstract}

\section{Motivation}

The insatiable demand for computational power for AI applications
  has resulted in a new breed of AI supercomputers
  that critically rely on advanced, high-bandwidth photonic interconnect fabrics
  for their operation~\cite{jouppi2023tpu}.
As integrated photonics embeds itself further and further into the core of 
  the modern computer hardware stack~\cite{lu2024high},
  the availability of fast, scalable photonic simulation engines 
  to accelerate and even revolutionize photonic simulation, design, and verification workflows
  becomes increasingly necessary to continue the pace of innovation~\cite{wade2020teraphy}.

A high-throughput, cost-effective photonic simulation engine could drive such a 
  ``revolution-within-a-revolution''---revolutionizing photonics to, in turn,
  continue to power the AI revolution---by removing
  the expensive, months-long development cycle of multi-project wafer runs and
  laboratory-based device characterization and replacing it with high-fidelity
  simulation-driven prediction of real-world performance,
  as well as enabling new paradigms like inverse design~\cite{piggott2017fabrication}
  which leverage high-throughput simulation capabilities
  to automatically optimize existing structures in a free-form manner,
  or else to search for better device topologies across vast design spaces.

Unfortunately,
  the mainstay and workhorse of photonic simulation algorithms,
  the finite-difference time-domain (FDTD) method~\cite{yee1966numerical},
  because it is a memory-intensive, but computationally-lightweight algorithm,
  is fundamentally misaligned with modern computational platforms,
  which are, instead, equipped with massive computational throughput
  and relatively meager memory bandwidth.

Here, we present an implementation of the original FDTD method
  that circumvents these limitations by use of a ``systolic'' update scheme,
  which relaxes both synchronization and memory bandwidth requirements,
  while still performing the exact same update equations.
Our implementation achieves a performance of 
  $\sim 0.18$ trillion cell updates per second (TCUPS)
  on a single Nvidia H100 GPU.
More importantly,
  we show that it largely regains the performance lost due to simulation domains
  exceeding the capacity of the top-level caches in the GPU memory hierarchy.

Our work paves the way for increasingly efficient implementations of the FDTD method
  on modern computational platforms,
  and even the design of new computational platforms
  custom-tailored to the systolic implementation of the FDTD method,
  as part of the mission-critical toolset needed to continue powering
  digital communication and computing in the AI era.

\section{Inherent mismatch between FDTD and GPUs}

The FDTD method presents a straightforward method 
  for updating Maxwell's equations in the time domain
  and consists of update equations in the vein of

\begin{equation}
  E_x^{i, j, k} = 
    E_x^{i, j, k} + 
      \frac{\Delta t}{\Delta y} \left( H_z^{i, j, k} - H_z^{i, j - 1, k} \right) -
      \frac{\Delta t}{\Delta z} \left( H_y^{i, j, k} - H_y^{i, j, k - 1} \right),
\end{equation}

which demonstrates that the update of a $E_x$ field value requires

\begin{itemize}
  \item 6 memory operations:
  \begin{itemize}
    \item 1 load of $E_x$,
    \item 2 loads of $H_z$,
    \item 2 loads of $H_y$, and
    \item 1 store of $E_x$.
  \end{itemize}
  \item 6 floating-point operations:
  \begin{itemize}
    \item 4 additions/subtractions, and 
    \item 2 multiplications
      (assuming $\Delta t/\Delta y$ and $\Delta t/\Delta z$ have been precomputed).
  \end{itemize}
\end{itemize}
If we assume 4-byte floating-point numbers,
  this simplified analysis reveals a compute-to-memory ratio of
  0.25 floating-point operations per byte of memory transferred.

In contrast,
  modern compute platforms such as the Nvidia H100 NVL~\cite{nvidia2024h100}
  which boasts $60 \times 10^9$ floating-point operations per second but a
  bandwidth of only $\sim 4 \times 10^9$ bytes per second,
  for a compute-to-memory ratio of $\sim 15$
  (note that if we assume tensor-core operations,
   this ratio increases yet further, by over an order of magnitude).
Herein lies the fundamental mismatch between the FDTD method and modern compute platforms,
  that the FDTD method is a computational workload
  that is nearly two orders of magnitude more memory-intensive
  than those for which modern compute platforms are optimized.
Or, put another way,
  that a naive implementation of the FDTD method
  is expected to unlock only $\sim 1\%$ of the computational capability
  of modern compute platforms.

That said, we extend our simple analysis by examining 
  the performance of the FDTD method 
  as the size of the simulation domain is increased.
To do this, we implement the FDTD method in a very straightforward manner,
  and observe that the performance
\begin{itemize}
  \item bottoms out at a per-thread performance of $\sim 0.2$ cell updates per microsecond,
    or $\sim 0.026$ TCUPS (128K threads in use)
    as the simulation domain increases,
    which is consistent with prior art~\cite{minkov2024gpu};
  \item but also has the critical characteristic
    of actually increasing dramatically by a factor of $\sim 10$
    for much smaller simulations.
\end{itemize}

\begin{figure*}[bp]
		\begin{center}
			\begin{pspicture}(-2,-2)(12,5) % Define the plotting area
    \psaxes[
        Dx=1,
        Dy=0.5,
        labels=y
    ]{->}(2,0)(2,0)(10,3.5) % Draw axes with logarithmic x

    \psline[linewidth=1.5pt]
    (2.885,2.710) (3.186,2.679) (3.487,2.679) (3.788,2.557) (4.089,2.542) (4.391,2.137)
    (4.391,2.137) (4.692,1.347) (4.993,0.734) (5.294,0.679) (5.595,0.521) (5.896,0.511)
    (6.197,0.506) (6.498,0.487) (6.799,0.467) (7.100,0.426) (7.401,0.326) (7.702,0.257)
    (8.003,0.238) (8.304,0.225) (8.605,0.225) (8.906,0.218)

    \psdots[dotsize=4pt]
    (2.885,2.710) (3.186,2.679) (3.487,2.679) (3.788,2.557) (4.089,2.542) (4.391,2.137)
    (4.391,2.137) (4.692,1.347) (4.993,0.734) (5.294,0.679) (5.595,0.521) (5.896,0.511)
    (6.197,0.506) (6.498,0.487) (6.799,0.467) (7.100,0.426) (7.401,0.326) (7.702,0.257)
    (8.003,0.238) (8.304,0.225) (8.605,0.225) (8.906,0.218)

    \psline[linestyle=dotted,linewidth=1pt](4.391,0)(4.391,3.5)
    \uput[0](4.391,3.0){global sync required}

    \psline[linestyle=dotted,linewidth=1pt](5.368,0)(5.368,2.5)
    \uput[0](5.368,2.0){L1 cache limit}

    \psline[linestyle=dotted,linewidth=1pt](7.720,0)(7.720,1.5)
    \uput[0](7.720,1.0){L2 cache limit}

    % Add custom labels
    \uput[d](2,-0.1){$10^2$}
    \uput[d](3,-0.1){$10^3$}
    \uput[d](4,-0.1){$10^4$}
    \uput[d](5,-0.1){$10^5$}
    \uput[d](6,-0.1){$10^6$}
    \uput[d](7,-0.1){$10^7$}
    \uput[d](8,-0.1){$10^8$}
    \uput[d](9,-0.1){$10^9$}

    \uput[d](6,-0.75){simulation size (bytes)}

    \uput[u](0.4,1.5){\rput{90}{
      \parbox{5cm}{\centering per-thread throughput \\ (updates/usec/thread)}
      }}
    % \uput[u](1, 3){\parbox{5cm}{\centering per-thread throughput \\ (updates/usec/thread)}}
\end{pspicture}
      \caption{\textbf{
  Empirical performance of a naive FDTD implementation
  for increasingly large simulation domains.}
  The performance of a naive implementation of the FDTD update algorithm
  on a H100 Nvidia GPU is shown for increasingly large simulation domains
  on a per-thread basis. At a high-level, our results show that there is an
  order-of-magnitude loss in performance that is predominantly attributable
  to the onset of the need to perform global synchronization operations,
  and to access memory resources which exceed first the capacity of the L1 cache,
  and then the capacity of the L2 cache, until finally the global memory pool is accessed.
  The bottomed-out performance at large simulation domains is congruent with previous
  work reported in the literature~\cite{minkov2024gpu}. At the same time,
  our results in this figure hint that better performance may be possible
  to be achieved by devising an update strategy which can capture the performance
  attainable when the simulation domain is so miniscule that it can fit entirely
  within the L1 cache of the GPU, and somehow extend that performance to arbitrary
  simulation domain sizes.
\label{fig:ref-perf}}
		\end{center}
	\end{figure*}
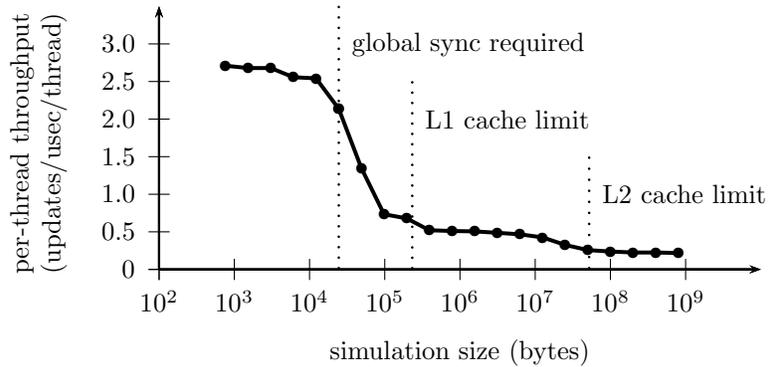

This order-of-magnitude performance gap can be attributed to three primary factors,
  as denoted in figure~\ref{fig:ref-perf}.
\begin{enumerate}
  \item the need for global synchronization when the number of threads
    used to update the simulation domain exceeds 1024,
    so that multiple thread blocks must now be used,
  \item the L1 cache limit of 228 KB on the H100 GPU, and
  \item the L2 cache limit of 50 MB on the H100 GPU.
\end{enumerate}

We conclude, then, that although there is a large, fundamental
  compute-to-memory ratio mismatch between the FDTD method
  and modern compute platforms,
  this mismatch is partially alleviated for very small simulation domains
  that both fit within a high-throughput, low-latency cache that is
  located near the arithmetic units of the processor
  (such as the L1 cache on the H100 GPU),
  and can bypass the need for the synchronization of threads
  at a global, chip-level scale.

\section{Systolic FDTD update scheme}

We now leverage the insight from our previous experiment
  and adopt the following strategy for circumventing the memory bandwidth
  bandwidth limitations of a GPU-accelerated FDTD engine.
Specifically, we propose that,
  instead of seeking a fast FDTD update over a single, monolithic simulation domain,
  we should instead seek to implement the FDTD update over a series of 
  \emph{smaller, interconnected simulation subdomains}.
We would then seek to circumvent the order-of-magnitude performance gap
  seen in figure~\ref{fig:ref-perf} by designing subdomains
  that both fit within the L1 cache of the GPU,
  and minimize the need of global synchronization between neighboring subdomains.

We term this strategy the \emph{systolic FDTD update scheme},
  since it is reminiscent of the systolic array architecture,
  in that each subdomain can be considered to be a single ``node''
  which communicates, as we shall see, only boundary values 
  to neighboring nodes, 
  and communicates them in a directional (as opposed to bi-directional)
  manner which minimizes the need for global synchronization.

While our strategy is easily stated,
  the interdependencies of the FDTD update equations
  do not, at first glance, naturally yield to such a strategy.
Because the full FDTD update in three spatial dimensions
  is actually an acyclic dependency graph in four dimensions
  (three spatial and one temporal dimension),
  we instead illustrate the principles of our strategy
  for the case where there is a single spatial dimension
  (and two dimensions in total).

\begin{figure*}[bp]
		\begin{center}
			\begin{pspicture}(0,-0.5)(6,6)
  \input{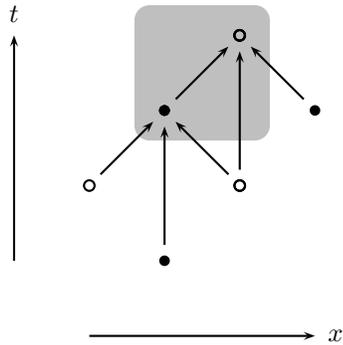}
	% Box around a single Yee cell.
	\psframe*[linecolor=lightgray,fillcolor=lightgray](1.6,2.6)(3.4,4.4)
	% Dots and arrows.
	\Hright12\Eright23
	\Eup21\Hup32
	\Hleft32\Eleft43
	% Axes.
	\Taxis014 \Xaxis140  
\end{pspicture}
      \caption{\textbf{
  Dependency structure for a single Yee cell in the 1D FDTD algorithm.}
  The dependency structure of a one-dimensional FDTD update algorithm
    is shown in the $x-t$ plane, for a single Yee cell where
    $E$-field nodes are represented by solid circles and
    $H$-field nodes are represented by hollow circles.
  The arrows indicate dependencies between field values;
    namely, that to compute the value of any given node,
    we must first obtain the values of all the nodes which point to it.
  As can be seen in the figure, each node is dependent
    on the value of the node at the previous timestep,
    as well as the adjacent nodes of the complementary field
    immediately adjacent to it.
\label{fig:single-yee-cell}}
		\end{center}
	\end{figure*}

For a single spatial dimension,
  the dependency structure of the FDTD update
  can be simplified to a scalar electric field
  interacting with a scalar magnetic field,
  where both are shifted by half a unit cell in both 
  spatial and temporal dimensions.
In this case, a single Yee cell
  can be considered as an $E$-field node paired
  with a neighboring $H$-field node
  as shown in figure~\ref{fig:single-yee-cell}.

\begin{figure*}[bp]
		\begin{center}
			\begin{pspicture}(0,-.5)(7,7)
  \input{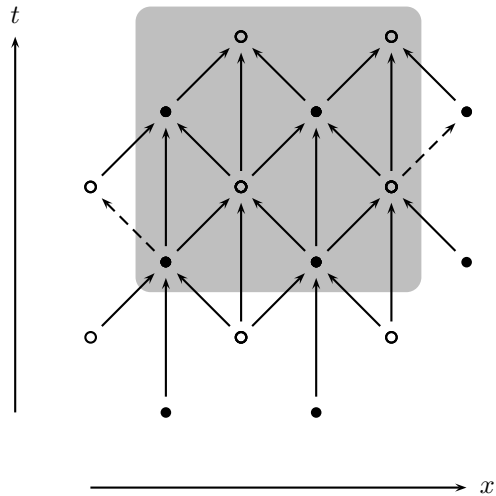}
	% Box around 2x2 cell.
	\psframe*[linecolor=lightgray,fillcolor=lightgray]
					(1.6,2.6)(5.4,6.4)
	% Dots and arrows.
	\multiput(1,0)(2,0){2}{
		\multiput(0,1)(0,2){2}{
			\Hright01\Eright12
			\Hleft21\Eleft32
			\Eup10\Hup21
		}
	}
	% Dashed arrows.
	\psset{linestyle=dashed}
  \Eleft23\Hright54  
	% Axes.
	\psset{linestyle=solid}
	\Taxis016\Xaxis160  
\end{pspicture}
      \caption{\textbf{
  $2 \times 2$ block of Yee cells as a computational subdomain.}
  A $M \times N$ block of Yee cells is the most natural
    subdivision of the FDTD computational domain
    but immediately reveals the problem of circular dependencies
    which arise at the block boundaries.
  This figure highlights the deadlocking condition found in a
    $2 \times 2$ block of Yee cells,
    in that the dashed arrows indicate nodes which are dependent
    on values within the computational subdomain,
    which the subdomain itself is dependent on.
  The deadlock condition generated by the circular dependencies
    means that such a tiling of the computational domain cannot
    be implemented in practice.\label{fig:2x2-cell}}
		\end{center}
	\end{figure*}

In our quest to device our systolic FDTD update scheme,
  we first attempt the most natural subdivision strategy,
  which is to simply divide the simulation domain both spatially and temporally;
  spatially, into groups of adjacent Yee cells,
  and temporally, by updating each spatial block
  independently of the others,
  deferring inter-block synchronization and communication 
  to later time steps.
This strategy is illustrated in figure~\ref{fig:2x2-cell},
  which shows an implementation which utilizes spatio-temporal blocks
  of shape $2 \times 2$.

\begin{figure*}[bp]
		\begin{center}
			\begin{pspicture}(0,-.5)(13,7)
  \input{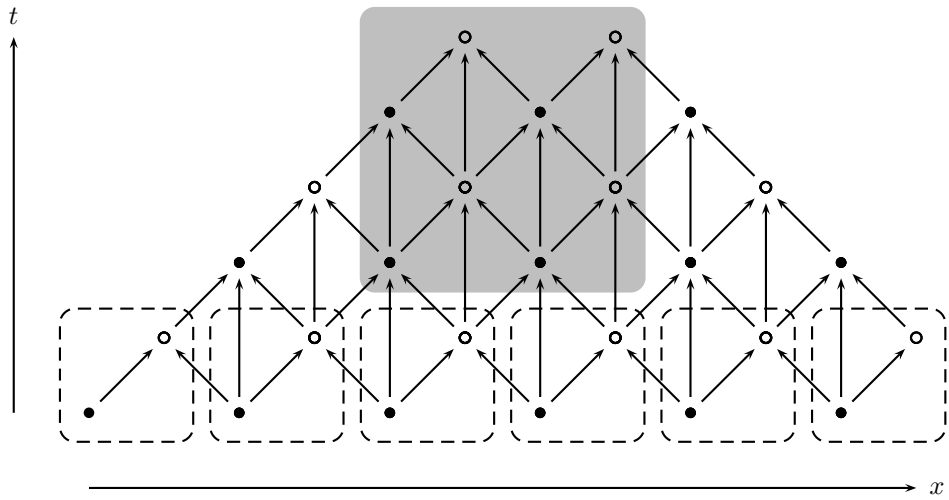}
	% Box around 2x2 cell.
	\psframe*[linecolor=lightgray,fillcolor=lightgray]
					(4.6,2.6)(8.4,6.4)
	% Required Yee cells.
	\multiput(0,0)(2,0){6}{
		\Eright11
		\psframe[linestyle=dashed](.6,.6)(2.4,2.4)
	}
	% Rest of dots and arrows.
	\multiput(3,1)(2,0){5}{\Eleft00}
	\multiput(2,1)(2,0){5}{\Hright01\Eup10\Hleft21}
	\multiput(3,2)(2,0){4}{\Eright01\Hup10\Eleft21}
	\multiput(4,3)(2,0){3}{\Hright01\Eup10\Hleft21}
	\multiput(5,4)(2,0){2}{\Eright01\Hup10\Eleft21}
	% Axes.
	\Taxis016\Xaxis{1}{12}{0}  
\end{pspicture}
      \caption{\textbf{
  Utilization of halo regions with a $2 \times 2$ block subdomain.}
	A $2\times2$ block updated exclusively
    from the field values of previous timesteps
    sidesteps the circular dependencies problem encountered in figure~\ref{fig:2x2-cell}
    at the cost of incurring the penalty of increased memory loads
    (as denoted by the nodes in the dashed boxes)
    and redundant computation in the overlapping regions of adjacent subdomains
    (as denoted by the nodes outside of the both the dashed boxes
    and outside of the $2 \times 2$ subdomain block).
\label{fig:2x2-halo}}
		\end{center}
	\end{figure*}

Unfortunately, figure~\ref{fig:2x2-cell} also reveals
  that this naive strategy is not only inefficient,
  but also results in a circular dependency, or deadlock condition,
  between blocks neighboring blocks.
The deadlock condition means that a practical implementation 
  of this strategy actually requires the use of ``halos'',
  as depicted in figure~\ref{fig:2x2-halo}.
Halo regions break the deadlocking condition,
  but come at the cost of increased memory usage,
  as well as the need to perform extra redundant computation.

\begin{figure*}[bp]
		\begin{center}
			\begin{pspicture}(0,-0.5)(8,8)
  \input{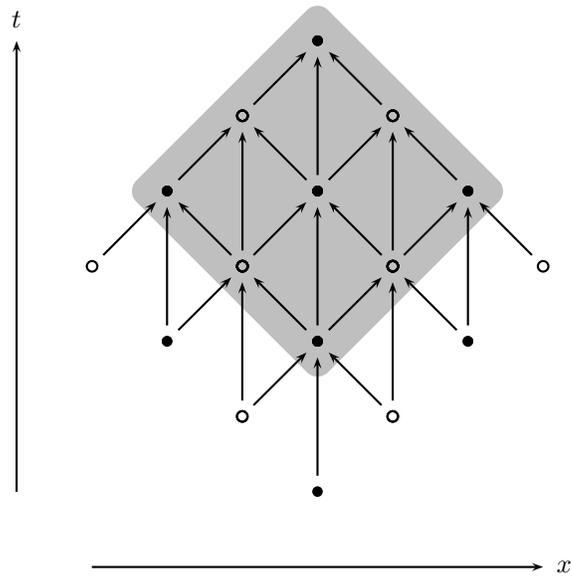}
	% Gray diamond.
	\pspolygon*[linecolor=lightgray,fillcolor=lightgray]
						 (4,2.45)(6.55,5)(4,7.55)(1.45,5)
	% Arrows and dots.
	\put(0,1){
		\Eup40
		\Hup31\Hright31\Hleft51\Hup51
		\Eup22\Eright22\Eleft42\Eup42\Eright42\Eleft62\Eup62
	}
	\multiput(1,4)(2,0){3}{\Hright00\Hleft20}
	\multiput(2,4)(2,0){2}{\Eright01\Hup10\Eleft21}
	\put(3,5){\Hright01\Eup10\Hleft21}
  % Axes.
	\Taxis017\Xaxis170
\end{pspicture}
      \caption{\textbf{
  The ``diamond''-shaped subdomain.}
	A diamond-shaped subdomain successfully avoids the deadlock conditions
    between adjacent subdomains encountered in figure~\ref{fig:2x2-cell},
    while also not requiring the use of halo regions
    and their associated overhead, as seen in figure~\ref{fig:2x2-halo}.
  The virtue of the diamond-shaped subdomain is that it is a natural
    subdivision of the computational domain
    in that it mimics the underlying dependency structure of the domain itself.
  In fact, the resulting inter-subdomain dependency structure 
    that arises when the simulation domain is decomposed into diamonds
    is identical to the original dependency structure of the FDTD update.
\label{fig:block-diamond}}
		\end{center}
	\end{figure*}

An alternative strategy is to break the deadlock condition
  without the cost of redundant computation or memory usage,
  comes in the form of the ``diamond'' block,
  which is a 45-degree rotation of the computational cell
  as shown in figure~\ref{fig:block-diamond}.
Such a strategy,
  by virtue of more naturally aligning with the dependency structure
  of the FDTD update,
  allows for computational subdomains which are deadlock-free,
  because each diamond effectively acts as a ``meta-node''
  in that the inter-diamond dependency structure is exactly the same
  as the original FDTD dependency structure---critically,
  each are only composed of one-directional dependencies.

\begin{figure*}[bp]
		\begin{center}
			\begin{pspicture}(0,-.5)(7,7)
	\input{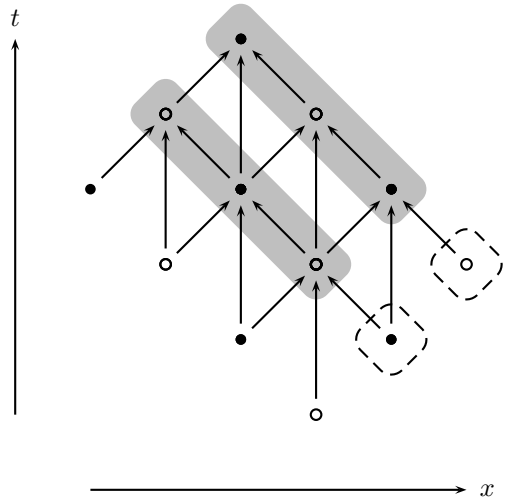}
	% Lassos.
	\pspolygon*[linecolor=lightgray,fillcolor=lightgray]
						 (4,2.45)(4.55,3)(2,5.55)(1.45,5)
	\pspolygon[linecolor=black,linestyle=dashed]
						 (4.45,2)(5,1.45)(5.55,2)(5,2.55)
	\pspolygon*[linecolor=lightgray,fillcolor=lightgray]
						 (5,3.45)(5.55,4)(3,6.55)(2.45,6)
	\pspolygon[linecolor=black,linestyle=dashed]
						 (5.45,3)(6,2.45)(6.55,3)(6,3.55)
	% Arrows and dots.
	\Eright14\Hup23\Hright23\Eup32\Eright32\Hup41
	\Hright25\Eleft34\Eup34\Eright34\Hleft43\Hup43\Hright43\Eleft52\Eup52
	\Hleft45\Eleft54\Hleft63
  % Axes.
  \Taxis016\Xaxis160
\end{pspicture}
      \caption{\textbf{
  A diagonal ``bar'' scanned along $x$-$t$.}
  Since the value of nodes at the same spatial location
    but at past time steps are not needed in general,
    the flattening of the diamond-shaped subdomain
    depicted in figure~\ref{fig:block-diamond}
    is a natural improvement.
  Such a bar-shaped subdomain is scanned along the $x$-$t$ diagonal
    because all the nodes which are needed to resolve
    the nodes at the new scan point are already present
    as nodes in the previous scan point, with the exception
    of a single extra node which must be loaded from global memory.
  As such, the diagonal bar subdomain represents a critical
    step towards a systolic update scheme.
  This is because it is the first subdomain which, by virtue of scanning,
    requires a lower-dimensional (zero dimensional point, in this case)
    shape to be loaded in order to resolve the nodes
    of a higher-dimensional (in this case, one-dimensional) shape.
  As such, it provides a critical stepping stone in achieving a systolic update scheme
    where communication between subdomains via global memory
    is necessary for boundary values of the subdomain only.
\label{fig:diagonal-drag}}
		\end{center}
	\end{figure*}

We can further improve the performance of the diamond block
  by scanning the diamond along the $x$-$t$ diagonal,
  as shown in figure~\ref{fig:diagonal-drag}.
This strategy is a natural extension of the diamond block,
  which naturally demands to be flattened along the temporal dimension
  (since there is no advantage in keeping track of field
  values for past time steps!),
  and gives us a first glimpse of how a systolic FDTD update scheme
  can be implemented.
This is because this is the first strategy which we have seen
  that exhibits the key property of being able to communicate
  boundary values at reduced dimensionality (in this case,
  a single zero-dimensional node)
  while continuing to update the interior of a computational cell
  at higher dimensionality (in this case,
  a one-dimensional diagonal bar).

\begin{figure*}[bp]
		\begin{center}
			\begin{pspicture}(0,-.5)(11,7)
	\input{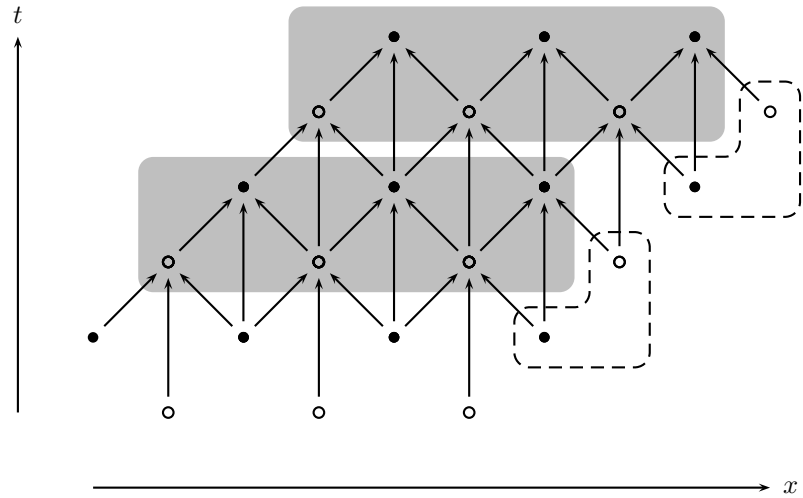}
  \multiput(0,0)(2,2){2}{
	  % Lassos.
	  \psframe*[linecolor=lightgray,fillcolor=lightgray](1.6,2.6)(7.4,4.4)
	  \pspolygon[linecolor=black,linestyle=dashed]
              (6.6,2.4)(6.6,1.6)(8.4,1.6)(8.4,3.4)(7.6,3.4)(7.6,2.4)
	  % Arrows and dots.
    \multiput(1,1)(2,0){3}{\Eright01\Hup10\Eleft21}
    \multiput(2,2)(2,0){3}{\Hright01\Eup10\Hleft21}
  }
  % Axes.
  \Taxis016\Xaxis1{10}0
\end{pspicture}
      \caption{\textbf{
  Horizontal-bar subdomain scanned along $x$-$t$.}
  The same diagonal scanning technique introduced in figure~\ref{fig:diagonal-drag}
    can also be applied to a horizontal bar of $N$ Yee cells.
  This results in an ideal subdomain scheme which possesses
    all the advantages of the diagonal bar in figure~\ref{fig:diagonal-drag},
    with the added advantage of being able to trivially resolve
    the nodes within each scan point.
  This is because in the case of the diagonal bar,
    each node must be resolved in a sequential manner,
    beginning from the node at the earliest time step
    (because nearly each node in the diagonal bar is dependent on a node
     from an earlier time step which is in the diagonal bar
     subdomain and must first be resolved),
    while in the case of the horizontal bar,
    all the nodes within the domain can be resolved in parallel
    (that is, all the $E$-field nodes can be resolved simultaneously,
    after which all the $H$-field nodes can then be resolved together).
\label{fig:flat-drag}}
		\end{center}
	\end{figure*}

A modified and final version of the strategy
  keeps the dynamic of scanning along the $x$-$t$ diagonal,
  but allows for much simpler inner update 
  in the utilization of a horizontal one-dimensional bar,
  as shown in figure~\ref{fig:flat-drag}
  (where all the field values are located at the same time step
   and can all be updated in a simultaneous manner),
  instead of the diagonal bar of figure~\ref{fig:diagonal-drag}
  (where the update of field values must proceed sequentially,
   starting from the node at the earliest time step).
As such, we consider the strategy depicted in figure~\ref{fig:flat-drag}
  to be the ideal form of our systolic FDTD update scheme,
  in that
\begin{itemize}
  \item it allows for the update of the 
    (in this case, one-dimensional, but in our practical implementation, three-dimensional) 
    subdomains to be proceed entirely in a local cache while
    only requiring lower-dimensional
    (a point in this case, but a two-dimensional slice in our practical implementation)
    set of boundary values to be communicated to neighboring subdomains
    via (much slower) global memory, and
  \item it allows for delayed synchronization of neighboring subdomains
    since an ``upwind'' subdomain can leave behind a trail of boundary values
    for a ``downwind'' subdomain to pick up whenever it is ready to consume them.
\end{itemize}
Finally, we note that this dynamic is only possible because of the one-directional dependencies
  between adjacent subdomains,
  which is directly a consequence of 
  \emph{forcing the subdomains to traverse along the $x$-$t$ diagonal},
  thus creating a one-directional dependency structure.
The downwind-only flow of boundary values in our systolic FDTD update scheme
  is simply a consequence of continually pushing the subdomains upwind.

\section{Practical implementation on an Nvidia H100 GPU}

Having established the principles of our systolic FDTD update scheme
  for the case of one-dimensional FDTD,
  we extend our strategy into a practical implementation
  of three-dimensional FDTD on an Nvidia H100 GPU,
  where subdomains are now three-dimensional blocks
  which traverse along the the major axes of the (four-dimensional)
  simulation domain, that is, along the $x-y-z-t$ diagonal.
Communication between subdomains now occurs in the form
  of two-dimensional slices, or faces, of the subdomains,
  which are stored in global memory,
  while the interior of the subdomains reside completely in L1 cache.

The details of our implementation are listed here:
\begin{itemize}
  \item Subdomains are formed from $31 \times 15 \times 15$ Yee cells,
    although a $32 \times 16 \times 16$ block of Yee cells is stored in cache,
    to allow for loading of boundary values from global memory.
  \item Each iteration involves the update of $31 \times 15 \times 15$ Yee cells,
    and the loading and writing of
    $\sim 32 \times 16 + 32 \times 16 + 16 \times 16$ cells to global memory
    (one slice or face for each of the three spatial dimensions).
  \item Compared to an implementation which loads, updates, and stores 
    exactly one Yee cell per iteration,
    this implementation requires $\sim 5.45$ times fewer global memory operations.
  \item Compared to the naive implementation
    consisting of a compute-to-memory ratio of $\sim 0.25$
    as depicted earlier, this implementation, in our estimation, achieves a
    compute-to-memory ratio of $\sim 4$
    which is over an order of magnitude improvement,
    but still far from being a compute-limited implementation.
  \item Our implementation doubles the number of memory operations needed per iteration,
    because it loads and stores not only electric- and magnetic-field values,
    but also structural fields corresponding to permittivity and conductivity
    structural fields (our implementation does not use PMLs, but 
    instead relies on the imaginary part of the permittivity to allow
    for the formation of adiabatic absorbers~\cite{oskooi2008failure}).
  \item On the other hand, our implementation also halves the number of bytes
    needed to be loaded per field value, by utilizing 16-bit floating-point 
    representations for the field values.
\end{itemize}

Because the performance of an FDTD simulation engine does indeed depend
  on simulation-specific details such as the size of the simulation domain,
  the number of timesteps, the source profile, and the output volume and format,
  we choose to present the performance of a stripped-down version of our
  engine, in line with what others have done in the past~\cite{minkov2024gpu}.
To this end, we choose to present the performance of our engine 
  for a ``bare-bones'' simulation setup, which is devoid of sources or output
  fields, but does include permittivity and conductivity fields.

For such a setup, we find that our engine achieves a performance of $\sim 0.15$ TCUPS.
This represents a $\sim 5.5$ times improvement over the performance of the naive implementation,
  as measured at large simulation domains.
In this sense, we have largely achieved our goal of circumventing the order-of-magnitude
  loss in performance due to the relatively slow memory bandwidth of the GPU
  and the need to perform global synchronization.

By devising a systolic update scheme which confines updates to smaller subdomains
  which individually fit within the L1 cache of the GPU,
  and by scanning these subdomains along the $x$-$y$-$z$-$t$ diagonal
  in order to minimize the frequency of global synchronization operations,
  we have been able to effectively extend the memory performance
  of the miniscule simulation residing only in the L1 cache of the GPU,
  to a large, practical-sized simulation domain only limited
  by the amount of global memory available to the GPU.

We attribute the remaining gap in performance to the fact that
  global synchronization operations are still required,
  that we have only been able to achieve partial overlap between
  the loading of boundary values from global memory and the update of the interior
  of the subdomain, and that there are still further optimizations 
  that can be made to our implementation.
We plan to make our implementation available as a simulation-as-a-service
  product that will be widely available to the public.

\section{Conclusion and future work}

The systolic FDTD update scheme presented here
  is a first step toward a future where high-throughput, large-scale
  photonic simulation capability is widely and generally available.
Such capabilities will likely come in the form of optimized computational
  kernels which continue to piggyback on the massive compute power
  of modern AI supercomputing centers (such as the Nvidia H100 GPU, as in our case),
  but may also evolve into a new generation of customized hardware
  which implement the systolic FDTD update scheme at the transistor-level;
  since in either case, the systolic FDTD update scheme will be pivotal
  in efficiently updating Maxwell's equations in the time domain.

We have motivated the need for such a systolic FDTD update scheme
  by showing both theoretically and empirically that the memory 
  bandwidth of compute systems is a very significant bottleneck
  to such a memory-intensive algorithm as FDTD,
  and we have shown how such a systolic update scheme
  can be used to circumvent the global memory bandwidth bottleneck
  of the GPU to enable a large-scale simulation of Maxwell's equations
  with performance mostly limited by the L1 cache bandwidth.

The future of photonics is bright,
  and as it continues to play an increasingly critical role in the continued development
  of the AI compute stack,
  we believe that the availability of increasingly powerful photonic simulation engines
  will also be an increasingly critical part of the photonic engineer's toolbox.
To that end, we have presented a general strategy as well as a practical implementation
  for performing large-scale FDTD simulations without being limited by the memory bandwidth
  of the global memory pool that such large-scale simulations must necessarily draw from.

\bibliographystyle{plain}
\bibliography{references}

\end{document}